# High-resolution gamma spectroscopy shift-invariant wavelet de-noising

Zhang Jinzhao[1](张金钊), Tuo Xianguo[1,2] （庹先国）

[1]Chengdu University of Technology Applied Nuclear Techniques in Geoscience Key Laboratory of Sichuan Province, Chengdu 610059, China

[2] Chengdu University of Technology State Key Laboratory of Geohazard Prevention & Geoenvironmental Protection, Chengdu 610059, China

**Abstract:** For the multi-resolution of wavelet transform, it used to filter the complex γ Spectrum, the fluctuation is filtered, while the detector resolution is still kept well, which has been demonstrated as a new, promising technique for γ spectrum de-noising. However, both side of the peak where the data rapidly changing area, the reconstruction spectrum will be artificial fluctuation of pseudo-Gibbs, when de-noising high-resolution γ spectrum. To solve these problems, a novel shift-invariant wavelet de-noising algorithm is proposed to treat the γ spectrum which measured by HPGe detector of the segment γ scanning system. It has a high resolution, a short measuring time, severe statistical fluctuation and scattering characteristics. The original spectrum was cycle spinning, de-noising by soft threshold, reconstructed. And then it was reversed cycle spinning, while the result was averaged. The algorithm not only overcomes the pseudo-Gibbs in the high-resolution γ spectrum de-noising by the traditional wavelet, but also keeps the shape of the characteristic peak well. Emphasis will be placed on quantifying the merits of this algorithm to traditional wavelet by exploring the quality of peak detection and localization.

**Keywords:** γ spectrum analysis, Wavelet transform, De-noising, Shift-invariant, Pseudo-Gibbs

## 1 Instruction

Segmented gamma scanning technology (SGS) is an important technical of the non-destructive nuclear waste qualitative and quantitative analysis. Nuclide qualitative and quantitative analysis of nuclear waste barrel can be reached by the HPGe detector of waste spontaneous and transmission gamma-ray measurement and the corresponding algorithm[1,2,3,4].

Gamma ray spectrum measurement is to know the information, the realization of the main methods for analysis of the radioactive nuclide in waste barrels. In the measurement, the radionuclide which was analyzed activity is low, low gamma ray energy, counting time is shorter, will cause the measured γ spectrum with a clear statistical fluctuation. To reduce the statistical fluctuation spectrum measurement, we usually use to extend the measurement time, or increase the strength of the source, as much as possible to improve the particle incident γ events. But in many cases, these two conditions are not easily achieved, then we need to use a certain algorithm for processing the measured spectra. While maintaining the shape of the spectrum, it maximum inhibition of γ spectrum of statistical fluctuations. The SGS system transmission gamma energy spectrum de-noise which based on the translation invariant wavelet in order to make more convenient accurate gamma energy spectrum analysis and provide the data basis for the qualitative and quantitative analysis nuclides.

Supported by National Natural Science Foundation of China(41274109, 41025015)

E-mail:zhangjinzhao_cdut@163.com

The commonly γ spectrum de-noising algorithm mainly gravity method, least squares smoothing method, Fourier transform, wavelet transform and discrete convolution function moves transform method. Among them, the least squares smoothing method and wavelet transform is most commonly used [5,6].

Least-square smoothing method has a good effect in the treatment of low energy resolution, more gentle when gamma energy spectrum data changes. but for high-resolution and changed more intense γ spectroscopy data, Its low-pass filter to filter out statistical fluctuations, but also filter out some of the really high-frequency signal to cause γ spectroscopy distortion. Wavelet transform processing low-resolution γ spectroscopy has better de-noising effect compared with the least squares filtering [7,8]. But when dealing with high-resolution γ spectroscopy, data transformation intense area where both side of the peak will appear for oscillation of pseudo Gibbs phenomenon. Based on the characteristics of wavelet transform, this study first proposed the use of shift invariant wavelet de-noising high-resolution γ spectroscopy, while maintaining the traditional wavelet transform spectroscopy at low resolution γ when good results, but also overcome the artificial data transformation intense region oscillation phenomenon.

## 2 Shift-invariant Wavelet transform

Shift-invariant wavelet transform is a de-noising algorithm based on wavelet transform. Wavelet transform theory: Signals are projected to a subspace which have different frequency, and then to process the signal in frequency space, finally the signal was reconstructed. Energy normalized wavelet family was obtained by basic wavelet function through stretch and translation: $\psi_{\tau,a}(t) = \dfrac{1}{\sqrt{a}} \psi(\dfrac{t-\tau}{a}) \quad a>0 \quad \tau \in R$ ,

$\psi_{\tau,a}(t)$ is a self-similar function family by scale $a$ extension transformation and time $\tau$ translation transformation of basic wavelet $\psi(t)$. We use this set of functions for analyzing signal decomposition [9].

Orthogonal wavelets are set of functions, which obtained the basic wavelet function stretching and translation. Translation is non-uniform sampling. With the increasing scale displacement sampling interval to 2 exponential times larger, orthogonal wavelet function set can not be from the perspective of multi-scale good match for the local structure of the signal characteristics. Therefore, the wavelet transform maybe generates oscillation in the signal drastic changes section [10,11,12].

Artificial oscillation phenomenon associated with signal singularity arrangement positions in the signal, and they have the same characteristics, but performance of a certain phase of the original signal, the new signal may produce smaller oscillation amplitude. Thus a effective way to eliminate artificial oscillation signal by changing the order of the arrangement, thereby changing the singularity position in the signal to reduce or eliminate the purpose of the oscillation amplitude. Thus there was a effective way to eliminate artificial oscillation signal by changing the order of the arrangement, the purpose is to change the singularity position in the signal to reduce or eliminate of the oscillation amplitude.

The steps of shift-invariant wavelet transform de-noising:

1. Shift.

First of all, the introduction of time-domain translation operator for signal $y_t$ ( $0 \le t \le N$ ) , and defined *Sh* as a shift operator which cyclic shift amount is h.

*Sh(xt)=x(t+h)modN*；

2. De-noising.

The signal with noise was transformed by wavelet, and then the wavelet function was de-noised with soft thresholding;

3. Opposite shift.

Opposite shift *S-h(xt)=x(t-h)modN*,

4. Average.

Artificially oscillation amplitude is minimized by selecting the optimal shift parameters h. However, when the spectrum contains a plurality of singular points, it is possible for a singular point shift amount is optimal, and the singular point to another shift amount is the worst. Therefore, for a complex spectrum, A range of shift was circulation shift operated, and then the result was averaged in order to eliminate the phenomenon of oscillation.

## 3 SGS Measurement Experiment

SGS system was carried out to the experiment which product by Canbera company. The system has a HPGe detector, a Collimator, shielding and other components. Measuring waste drums which filled Acrylonitrile Butadiene Styrene plastics（ABS） is the national standard 200L waste drums, drum wall thickness of 1.25mm. Sample volume: 5cm×5cm×5cm, density:1.07g/cm$^3$，radioactive source: $^{60}$Co，actively: 10μCi.

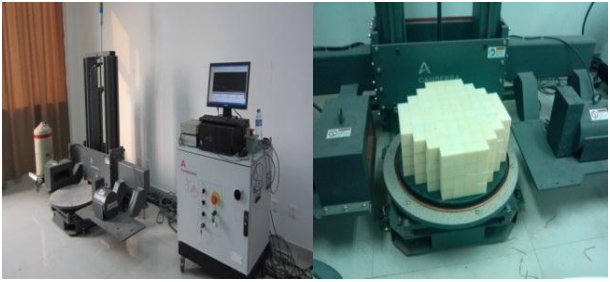

Fig. 1 SGS system schematic diagram

The energy spectrum which penetrated the empty drum, each piece containing 38 ABS plastic, and each piece containing 76 ABS plastic were measured by SGS system.

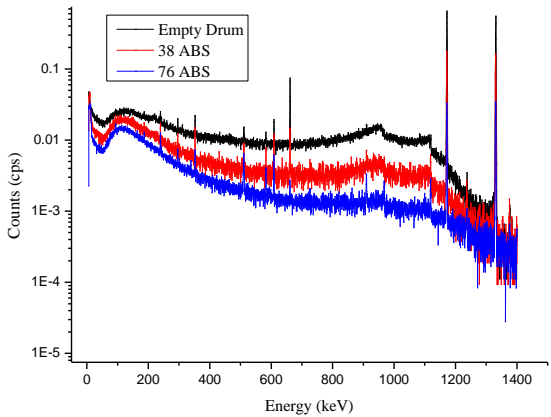

Fig. 2 The original γenergy spectrum of SGS system

The 1172keV and 1331keV peak is two full-energy peaks of $^{60}$Co.Fig.2 represent the original measurement γ energy spectrum of SGS system, as a function of the sample amount inside the waste drum. For all original γ energy spectrums have obvious statistical fluctuation, especially the high-energy, the statistical fluctuation increases with increasing sample amounts inside the waste drum.

## 4    Analysis and discussion

In order to reduce the original spectrum of statistical fluctuations, We take the empty drum original measurement energy spectrum as a  case to de-noise with five-point Least Square Filtering , eleven-point filtering Least Square Filtering, wavelet transform filtering, translation invariance wavelet transform filtering respectively.

Five point and eleven point least square filtering formula respectively：

$$\bar{Y}_i = \frac{1}{35}\left(-8y_{i-2}+12y_{i-1}+17y_i+12y_{i+1}-8y_{i+2}\right)$$

$$\bar{Y}_i = \frac{1}{429}(-36y_{i-5}+9y_{i-4}+44y_{i-3}+69y_{i-2}+84y_{i-1}+89y_i+84y_{i+1}+69y_{i+2}+44y_{i+3}+9y_{i+4}-36y_{i+5})$$

Peak position relative, counting the relative error and root mean square error (RMSE) as a measure of the noise reduction effect are smaller that de-noising spectrum deformation is less, and the de-noising effect is better. The best de-noising algorithms was determined by comparing the results.

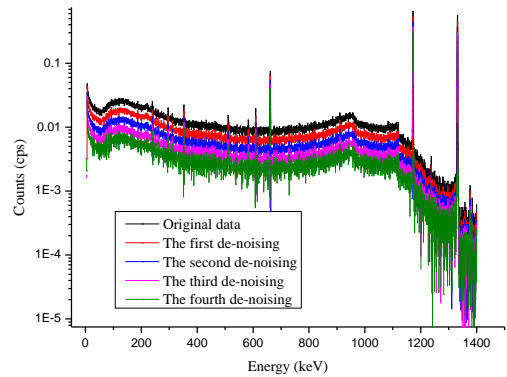

Fig. 3 Five-point least squares de-noising

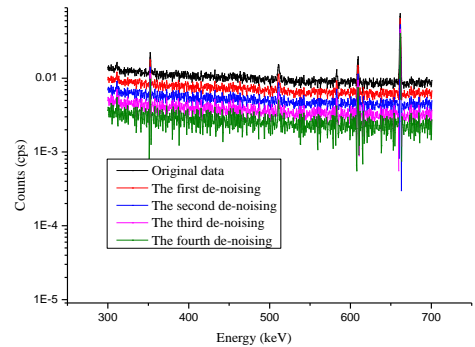

Fig. 4 Five-point least squares de-noising (300keV-700keV)

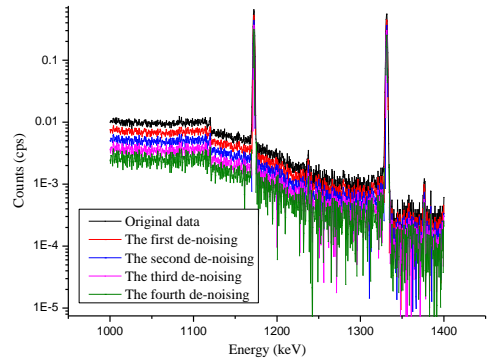

Fig. 5 Five-point least squares de-noising (1000keV-1400keV)

Table 1 Result of Five-point least squares de-noising

| Five-point LS de-noising | Peak position(keV) | Peak position relative error | Peak counts (cps) | Peak counts relative error | RMSE |
|---|---|---|---|---|---|
| Original data | 1172.29 | 0 | 0.65367 | 0 | 0 |
| | 1331.68 | 0 | 0.55694 | 0 | |
| First | 1172.29 | 0 | 0.53253 | 0.18532 | 0.007432 |
| | 1331.31 | $2.78 \times 10^{-4}$ | 0.44609 | 0.19903 | |
| Second | 1172.29 | 0 | 0.43948 | 0.32767 | 0.012854 |
| | 1331.68 | 0 | 0.36256 | 0.34901 | |
| Third | 1172.29 | 0 | 0.36944 | 0.43482 | 0.016879 |
| | 1331.31 | $2.78 \times 10^{-4}$ | 0.30047 | 0.46050 | |
| Fourth | 1172.29 | 0 | 0.31593 | 0.51668 | 0.019859 |
| | 1331.68 | 0 | 0.25251 | 0.54661 | |

Fig. 3 represents the four times de-noising result of five-point least square filtering. Fig. 4 and Fig. 5 is the enlarged display of Fig. 3 energy range 300keV-700keV and 1000keV-1400keV respectively. Table 1 represents the result of peak position and peak count rate, and the RMSE of four times de-noising. As shown in Fig. 3 and Table1, using five-point least squares filtering noise, Low-energy part of the spectrum deformation is small compared with the original data. High-energy part of the spectrum deformation is large compared with the original data, and the energy spectrum count rate decreased significantly. It will make the all-around peak area slants small to affect the measurement precision of the original. The results showed that γ spectroscopy is serious distortion after four times de-noising.

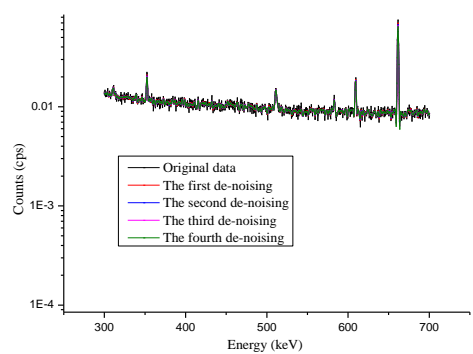

Fig. 7 Eleven -point least squares de-noising
(300keV-700keV)

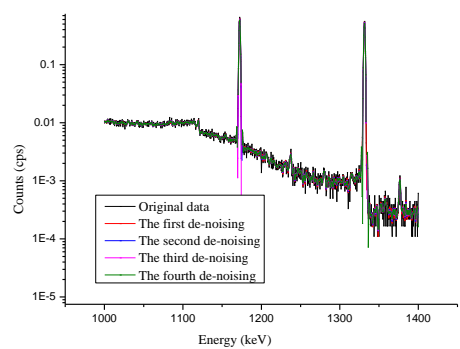

Fig. 8 Eleven -point least squares de-noising
(1000keV-1400keV)

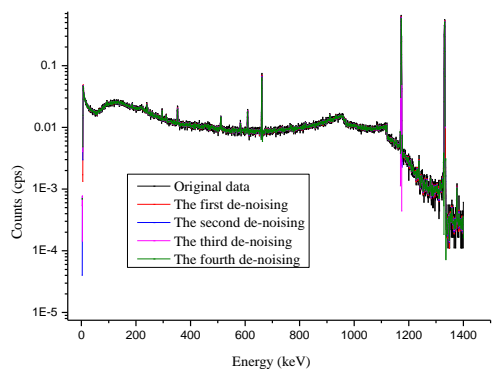

Fig. 6 Eleven-point least squares de-noising

Table 2 Result of Eleven-point least squares de-noising

| Eleven-point LS de-noising | Peak position(keV) | Peak position relative error | Peak counts (cps) | Peak counts relative error | RMSE |
|---|---|---|---|---|---|
| Original data | 1172.29 | 0 | 0.65367 | 0 | 0 |
| | 1331.68 | 0 | 0.55694 | 0 | |
| First | 1172.29 | 0 | 0.62393 | 0.04450 | 0.001312 |
| | 1331.31 | $2.78 \times 10^{-4}$ | 0.53828 | 0.03350 | |
| Second | 1172.29 | 0 | 0.60827 | 0.06945 | 0.001873 |
| | 1331.31 | $2.78 \times 10^{-4}$ | 0.52672 | 0.05426 | |
| Third | 1172.29 | 0 | 0.59595 | 0.08830 | 0.002355 |
| | 1331.31 | $2.78 \times 10^{-4}$ | 0.51818 | 0.06959 | |
| Fourth | 1172.29 | 0 | 0.58616 | 0.10328 | 0.002751 |
| | 1331.68 | $2.78 \times 10^{-4}$ | 0.51092 | 0.08263 | |

Fig. 3 represents the four times de-noising result of eleven-point least square filtering. Fig. 4 and Fig. 5 is the enlarged display of Fig. 3 energy range 300keV-700keV and 1000keV-1400keV respectively. Table 1 represents the result of peak position and peak count rate, and RMSE of four times de-noising. As shown in Fig. 3 and Table 2, using eleven-point least squares filtering de-noise, de-noised spectra is good agreement with the original data, but the peak count rate also decreased significantly affected the original measurement accuracy. The results showed that eleven-point least square filtering which maximum loss of counts is less 10% is better than five-point least square filtering for reducing peak counts and RMSE.

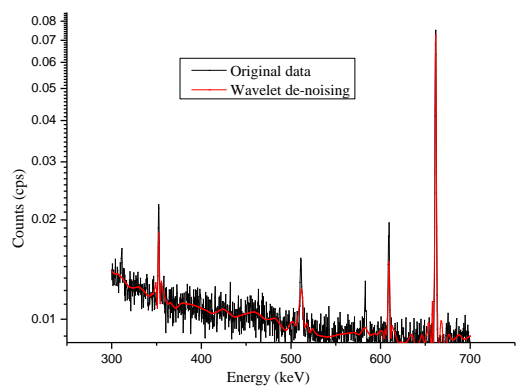

Fig. 10 Wavelet de-noising (300keV-700keV)

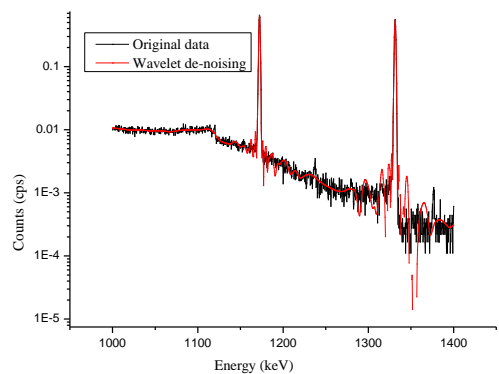

Fig. 11 Wavelet de-noising (1000keV-1400keV)

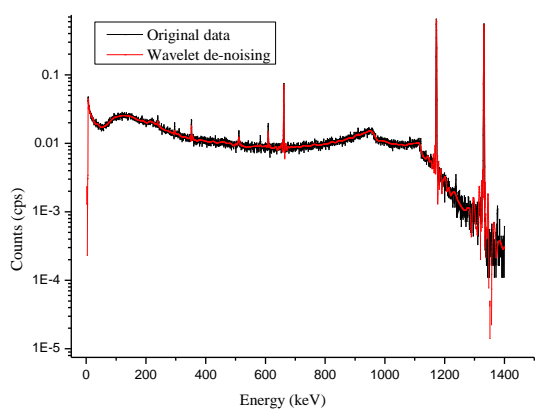

Fig. 9 Wavelet de-noising

Table 3 Result of wavelet de-noising

| Wavelet de-noising | Peak position(keV) | Peak position relative error | Peak counts (cps) | Peak counts relative error | RMSE |
|---|---|---|---|---|---|
| Original data | 1172.29 | 0 | 0.65367 | 0 | 0 |
| | 1331.68 | 0 | 0.55694 | 0 | |
| First | 1172.29 | 0 | 0.64292 | 0.16446 | 0.001012 |
| | 1331.31 | $2.78 \times 10^{-4}$ | 0.54767 | 0.01664 | |

Fig. 3 represents the de-noising result of wavelet transform filtering. Fig. 4 and Fig. 5 is the enlarged display of Fig. 3 energy range 300keV-700keV and 1000keV-1400keV respectively. As shown in Fig. 3, using wavelet transform filtering de-noise, de-noised spectra is good agreement with the original data, the peak count rate is not decreased significantly, but in the sides of the two

Table 3 represents that，At the flat part of data the wavelet transform filtering conformity with the original data better than the least squares, and peak counts less loss, RMSE significantly reduced, but both sides of the full-energy peaks have a clear artificial fluctuations. De-noising effect is better than least squares filtering.

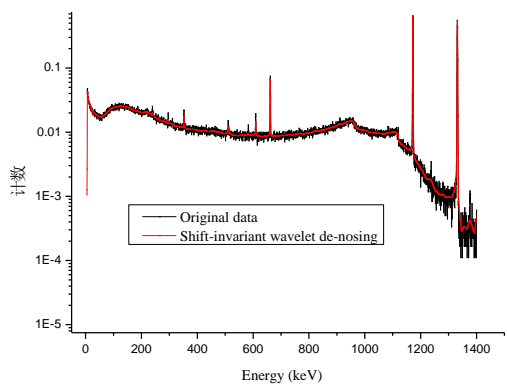

Fig. 12 Shift-invariant wavelet de-noising

full-energy peak fluctuation phenomenon is very obvious, affected the full spectrum of de-noising effect.

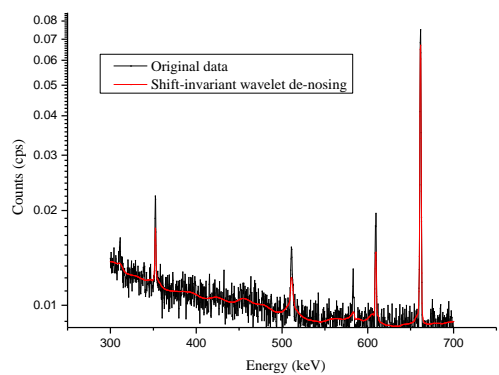

Fig. 13 Shift-invariant wavelet de-noising
(300keV-700keV)

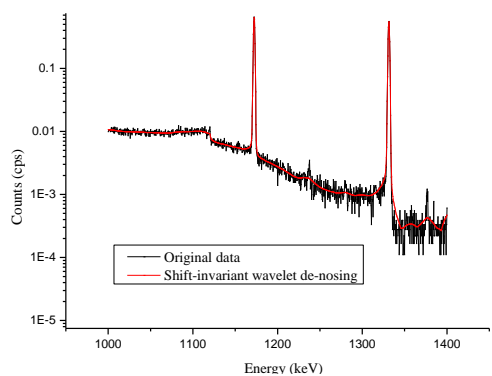

Fig. 14 Shift-invariant wavelet de-noising
(1000keV-1400keV)

Table 4 Result of Shift-invariant wavelet de-noising

| Shift-invariant wavelet de-noising | Peak position(keV) | Peak position relative error | Peak counts (cps) | Peak counts relative error | RMSE |
|---|---|---|---|---|---|
| Original data | 1172.29 | 0 | 0.65367 | 0 | 0 |
|  | 1331.68 | 0 | 0.55694 | 0 |  |
| First | 1172.29 | 0 | 0.64506 | 0.01317 | $8.74953 \times 10^{-4}$ |
|  | 1331.68 | 0 | 0.54970 | 0.01300 |  |

Fig. 3 represents the de-noising result of shift-invariant wavelet transform filtering. Fig. 4 and Fig. 5 is the enlarged display of Fig. 3 energy range 300keV-700keV and 1000keV-1400keV respectively. As shown in Fig. 3, using shift-invariant wavelet transform filtering de-noise, de-noised spectra is good agreement with the original data, the peak count rate is not decreased significantly, both sides of full-energy peak and the small peaks of low-energy part are very smooth, the de-noising effect is good. As shown in Fig. 3, shift-invariant wavelet transform filtering not only de-noise the energy spectrum of the Compton platform noising, reducing the statistical fluctuations, but also distinguish the 352keV peak of $^{214}$Pb, $^{214}$Bi peak of 609keV of environmental background and 511keV electron escape peak which improve the original the resolution of the measurement data, and also illustrate translation invariant wavelet transform has a weak peak detection capability. As shown in Fig. 4, both sides of full-energy peaks is very smooth, does not appear artificial oscillations, indicating that the algorithm can effectively inhibit the pseudo-Gibbs phenomenon.

Table 1 represents that, the relative error of shift-invariant wavelet transform filtering peak positions and peak count is far less than the least squares filtering and traditional wavelet transform. Shift-invariant wavelet transform filtering RMSE is minimum, indicating that shift-invariant wavelet transform filtering highest similarity with the original data and the data is most credible.

Fig. 15 is the de-noising spectroscopy which used shift-invariant wavelet transform filtering of Fig. 2 As shown in Fig.15, shift-invariant wavelet transform filtering de-noising achieved good results.

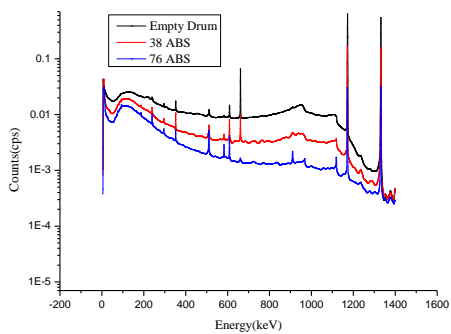

Fig. 15 The original γenergy spectrum of SGS system

## 5 Conclusion

In high-resolution gamma energy spectrum de-noising, the least squares filtering cannot eliminate the statistical fluctuation, and maintain the shape of the gamma energy spectrum. Traditional wavelet transform de-noising will appear Gibbs phenomenon where data changing larger portions on both sides of the peak, affecting de-noising effect. Since the total measurement time constraints of SGS measurement system, a smaller γ statistical fluctuation spectrum which got in a short time is required. The algorithm not only overcomes the pseudo-Gibbs in the high-resolution γ spectrum de-noising by the traditional wavelet, but also keeps the shape of the characteristic peak well.

## 6 Acknowledgements

The authors wish to acknowledge the support of the Chengdu University of technology Applied Nuclear Techniques in Geoscience Key and State Key Laboratory of Geohazard Prevention & Geoenvironmental Protection.